\documentclass{article}

\usepackage{arxiv}

\usepackage[utf8]{inputenc}
\usepackage[T1]{fontenc}
\usepackage{hyperref}
\usepackage{url}
\usepackage{booktabs}
\usepackage{amsfonts}
\usepackage{nicefrac}
\usepackage{microtype}
\usepackage{graphicx}
\usepackage{natbib}
\usepackage{doi}
\usepackage[USenglish]{babel}
\usepackage[intlimits]{amsmath}
\usepackage{bm}
\hypersetup{colorlinks=true, citecolor=blue}
\usepackage{amssymb}
\usepackage{fontawesome}
\usepackage{mathtools}
\usepackage{xcolor}
\usepackage{lmodern}
\usepackage{siunitx}
\usepackage{booktabs}
\usepackage{etoolbox}

\renewrobustcmd{\bfseries}{\fontseries{b}\selectfont}
\renewrobustcmd{\boldmath}{}
\newrobustcmd{\B}{\bfseries}

\usepackage{doi}
\usepackage{titlesec}

\titleformat{\paragraph}
{\normalfont\normalsize\bfseries}{\theparagraph}{1em}{}
\titlespacing*{\paragraph}
{0pt}{3.25ex plus 1ex minus .2ex}{1.5ex plus .2ex}

\title{A Review and Critique of Auxiliary Information-Based Process Monitoring Methods}

\date{September 30, 2021}

\author{ \href{https://orcid.org/0000-0002-3618-6746}{\includegraphics[scale=0.06]{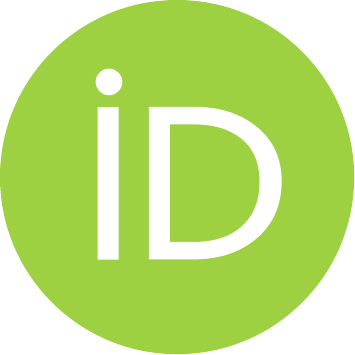}\hspace{1mm}Nesma A. Saleh} \\
	Dep. of Statistics\\
	Faculty of Economics and Political Science\\
	Cairo University \\
	\texttt{neasaleh@feps.edu.eg}\\
	\And 
	\href{https://scholar.cu.edu.eg/?q=mamahmou}{\faHome\hspace{1mm}Mahmoud A. Mahmoud} \\ 
	Dep. of Statistics\\
	Faculty of Economics and Political Science\\
	Cairo University \\
	\texttt{mamahmou@feps.edu.eg}\\
	\And
	\href{https://www.stat.vt.edu/people/stat-faculty/woodall-bill.html}{\faHome\hspace{1mm}William H. Woodall} \\
	Dep. of Statistics\\
	Virginia Tech\\
	Blacksburg VA, USA\\
	\texttt{bwoodall@vt.edu} \\
	\And
    \href{https://orcid.org/0000-0002-9666-5554}{\includegraphics[scale=0.06]{orcid.pdf}\hspace{1mm}Sven Knoth} \\
    Dep. of Mathematics \& Statistics\\
    Helmut Schmidt University\\
    Hamburg, Germany\\
    \texttt{knoth@hsu-hh.de} \\
}

\renewcommand{\shorttitle}{Review \& Critique of AIB-PM Methods}

\begin{document}
	
\maketitle
	
\begin{abstract}
We review the rapidly growing literature on auxiliary information-based (AIB) process monitoring methods. Under this approach, there is an assumption that the auxiliary variable, which is correlated with the quality variable of interest, has a known mean, or some other parameter, which cannot change over time. We demonstrate that violations of this assumption can have serious adverse effects both when the process is stable and when there has been a process shift. Some process shifts can become undetectable. We also show that the basic AIB approach is a special case of simple linear regression profile monitoring. The AIB charting techniques require strong assumptions. Based on our results, we warn against the use of AIB approach in quality control applications.
\end{abstract}


\section{Introduction} % \label{sec:intro}

Quality control charts are considered one of the effective tools of statistical process control/ monitoring (SPC/M). An appropriate use of control charts aims to detect unusual changes in the process norm. Generally, in practical situations, a process could be monitored using a single (univariate) or multiple (multivariate) quality characteristics (QCs). In univariate monitoring, the quality of a product/service is defined based on a sole component. In multivariate monitoring, however, the quality is defined based on two or more components. Accordingly, in multivariate quality control it is required to simultaneously monitor and control several QCs/properties.

\cite{Hote:1947} initiated the work on multivariate processes by introducing the Hotel\-ling’s $T^2$ chart;
which is considered a multivariate generalization of the Shewhart chart. The Shewhart chart was the first control chart, introduced in the SPC/M literature by \cite{Shew:1924, Shew:1925, Shew:1926}.
Both Shewhart and $T^2$ charts are known for their efficient detection of large changes in the process parameters. Later, the multivariate chart technique was extended to other control charts; e.g. the cumulative sum (CUSUM) chart
introduced by \cite{Page:1954c}, and the exponentially weighted moving average (EWMA) chart
introduced by \cite{Robe:1959}. Both charts are well-known for being sensitive to small and moderate magnitudes of process changes. The multivariate EWMA (MEWMA) and multivariate CUSUM (MCUSUM) charts are their corresponding multivariate
extensions \citep{Wood:Ncub:1985, Heal:1987, Cros:1988, Pign:Rung:1990, Lowr:Wood:Cham:Rigd:1992}.
Reviews of multivariate charts were conducted by
%Alt (1984), Jackson (1985), Lowry and Montgomery (1995), and Bersimis et al. (2007).
\cite{Jack:1985}, \cite{Lowr:Mont:1995}, \cite{Alt:2004} and \cite{Bers:Psar:Pana:2007}.

One of the main drawbacks of multivariate control charts is that one cannot always easily identify the source that caused the process to be out-of-control. Additionally, in some practical situations, the QC of interest (process output) could be influenced by some external factors. Conventional charts monitor the process output directly without taking into consideration these influential factors. Hence, other alternative techniques were developed to monitor multivariate processes rather than the standard multivariate charts. Among these techniques were the regression chart proposed by
%Mandel (1969),
\cite{Mand:1969},
regression-adjusted charts proposed by
%Hawkins (1991, 1993),
\cite{Hawk:1991, Hawk:1993a}
cause-selecting charts proposed by
%Zhang (1984, 1985, 1989, 1992)
\cite{Zhan:1984, Zhan:1985, Zhan:1990, Zhan:1992}
and
%Constable et al. (1987, 1988),
\cite{Cons:Clea:Zhan:1987, Cons:EtAl:1988}
and profile monitoring techniques studied by many authors including, for example,
%Kang and Albin (2000).
\cite{Kang:Albi:2000}.

In regression charts, the regression analysis is integrated with conventional control charts after a regression model is fitted to account for the effect of an external covariate on the process output. That is to say, the process variable is monitored after removing the influence of the covariate, and then charting the regression residuals. For example,
%Mandel (1969)
\cite{Mand:1969}
used the regression chart in monitoring the number of hours required to handle and sort mail (process output) based on their volume (external covariate). Regression charts are useful for monitoring a varying average instead of a constant one as conventional charts do, and one of its most important applications is monitoring and controlling multistage processes. 

The regression chart was extended to a regression-adjusted chart by plotting on univariate charts the residuals of each variable obtained when regressing this variable on all the other variables
%(Hawkins (1991)).
\citep{Hawk:1991}.
The aim of this extension was to maximize the sensitivity to a structured mean shift.
%Hawkins (1993)
\cite{Hawk:1993a}
extended the regression-adjusted scheme to be applicable for monitoring cascade (sequential value-added) processes. In a cascade process, the product quality at the end of a given step depends simultaneously not only on how well this given step was performed, but also on the outcomes of preceding steps. The quality after a particular process step must be adjusted based on the outcomes of previous steps before a meaningful decision can be made about the current step. The cause-selecting chart is based on a similar concept to that of
%Mandel’s (1969)
\citeauthor{Mand:1969}'s (\citeyear{Mand:1969})
regression chart of monitoring the process output after removing the influential input covariates.
%Wade and Woodall (1993)
\cite{Wade:Wood:1993}
provided an extensive review of cause-selecting charts.

More recently profile monitoring techniques have been developed in which a profile refers to a functional relationship between a dependent variable and one or more explanatory variables. Profile monitoring, in fact, can be considered as well an extension to the regression control chart. However, in the profile monitoring techniques the main objective is to monitor the stability of this relationship over the time; and hence one monitors the profile parameters with the aim to detect changes in the process
%(See Woodall et al. (2004); Woodall (2007a); Woodall (2007b); Noorossana et al. (2011); Maleki et al. (2018)).
\citep{Wood:EtAl:2004, Wood:2007, Wood:2008, Noor:Sagh:Amir:2011, Male:Amir:Cast:2018}.

Even more recently,
%Riaz (2008a, 2008b)
\cite{Riaz:2008a, Riaz:2008b}
proposed the use of some auxiliary information to enhance the precision of the estimator used in monitoring the parameters of the QCs of interest. In the survey sampling literature, auxiliary information is useful when it comes to estimating the unknown population parameters. It has been shown that auxiliary information increases the precision/efficiency of estimators when sampling from a fixed sampling frame
%(Cochran (1977); Naik and Gupta (1991); Singh and Mangat (1996)).
\citep{Coch:1977, Naik:Gupt:1991, Sing:Mang:1996b}.
Auxiliary variables are correlated to the variable of interest. The auxiliary variable is identified to be easy, available, and inexpensive to measure comparing to the variable of interest. A considerably growing number of researchers have proposed integrating the auxiliary information-based (AIB) estimators into control charts.
We review this literature in Section~\ref{sec:RAIB}. 

In the many studies, researchers have mentioned several different practical processes where the use of an AIB chart would be applicable. In general we do not find these examples convincing. In some cases the examples involve simply monitoring two or more quality variables. For example,
%Ahmad et al. (2014b)
\cite{Ahma:EtAl:2014b}
stated that monitoring the amount of filled-in liquid in some filling process could be supplemented by the easy-to-observe filling speed as an auxiliary characteristic. In this case the mean of the filling speed would have to be assumed to remain constant with no provision for monitoring it.
%Ahmad et al. (2013a)
\cite{Ahma:EtAl:2013}
gave examples on processes that would use more than one auxiliary characteristic; such as monitoring the amount of total power generated from coal that could be supplemented by both air temperature, humidity, and the amount of flue gas as easy-to-measure auxiliary characteristics. It is not reasonable to assume, however, that these auxiliary variables have a constant mean. Some examples of potential applications seem questionable and inappropriate for the AIB theory, such as monitoring the progression of a wave produced on electrocardiography (ECG) machine of a heart patient per minute supplemented by two auxiliary characteristics, blood pressure and cholesterol level. Further such examples can be found, for example, in
\cite{Riaz:Does:2009}, \cite{Riaz:EtAl:2013}, \cite{Haq:Khoo:2016} and \cite{Noor:Khan:Asla:2018}.
%Riaz and Does (2009), Riaz et al. (2013), Haq and Khoo (2016), and Noor-ul-Amin et al. (2018).

The assumption across all these studies is that the auxiliary characteristic(s) distributional properties are known and stable (remains unchangeable) over the monitoring time of the QC of interest.
%Hawkins (1993)
\cite{Hawk:1993a}
differentiated between two different mechanisms for correlated QCs of a certain product/service. The first mechanism is that a change (shift) in any of the characteristics does not affect the others, while the second mechanism is related to the characteristics with natural ordering by which a change in any of the characteristics affects some or all the following ones, but not the preceding ones (i.e., the cascade property). The proposed and presented AIB charts in literature are evaluated under the assumption of the first mechanism stated by
%Hawkins (1993),
\cite{Hawk:1993a},
i.e. processes without the cascade property
%(Riaz and Does (2009); Abbasi and Riaz (2013); Ahmad et al. (2014a)). 
\citep{Riaz:Does:2009, Abba:Riaz:2013, Ahma:EtAl:2014a}.

In our opinion, the assumed stability of the auxiliary characteristic(s) in quality monitoring and control is a very strong one. Based on the AIB research, the auxiliary information comes in the form of process characteristics that are related to the quality characteristic of interest. It is well-established in quality control applications that manufacturing processes are subject to assignable causes of variation. Hence, the assumption that the auxiliary characteristic(s) would remain stable during the monitoring time is unrealistic unless the auxiliary variable is under some type of engineering control.  Even then it would seem prudent to monitor the auxiliary variable, something not considered under the AIB framework.

In some cases it does not seem that AIB researchers understand the required assumptions.
%Ahmad et al. (2014a)
\cite{Ahma:EtAl:2014a}
wrote that the AIB method can be applied in the following situation:

\vspace*{1ex}
\begin{center}
\begin{minipage}{0.9\textwidth}
\textit{``A process where the study variable is correlated with one or more auxiliary variables but the study variable can endure a shift in location or dispersion parameter without affecting the auxiliary variables. Hawkins (1991) elaborated this kind of situation using the example of platinum refinery. Here the quality of the process may depend upon the quantity of platinum metal which is generally correlated with the magnitude of other metals. So it is quite possible for the workers to disturb the concentration of platinum metal in the process. However, other metals will not be affected by this shift in the study variable.''}
\end{minipage}
\end{center}
\vspace*{1ex}

\noindent
There is no quality variable in the application considered by
%Hawkins (1991),
\cite{Hawk:1991},
however, that can be assumed to have a constant mean. Thus none can be used as auxiliary variables.

Further, we show in our paper that a change in the mean of an auxiliary variable can increase the rate of false alarms and mask changes in the process characteristic of interest. 

One objective of our paper is to provide an extensive literature review on control charts proposed for integrating the AIB estimation technique. Additionally, we further discuss several important issues regarding the applicability of using AIB estimation in quality control monitoring. We also show a link between the AIB charts and profile monitoring methods. 

The paper is organized as follows:
In Section~\ref{sec:EuAI}
we present some of well-known and commonly used AIB estimators used in sampling applications.
In Section~\ref{sec:RAIB}
we provide an extensive literature review on the AIB control charts.
In Section~\ref{sec:AIBissues}
we present further arguments about the pitfalls of integrating the AIB-estimators in quality control applications. This is done through some derivations and simulated examples.
In Section~\ref{sec:AIBprofile}
we discuss the links between AIB charts and the profile monitoring approach. Finally,
Section~\ref{sec:discussion}
contains a brief discussion on our findings.

\section{Estimation Using Auxiliary Information} \label{sec:EuAI}

A considerable amount of research has been devoted to different techniques to integrate the auxiliary information/variables in estimating the unknown population parameters in sampling applications. It is well-known in the survey sampling literature that the use of auxiliary information reduces the variance of the estimators of the parameters of interest. We present in this section some of the commonly used AIB estimators. We let $Y$ represent the variable for which we are interested in estimating the mean ($\mu_Y$), and let $X$ represent an auxiliary variable that is correlated with $Y$. We assume having a simple random sample of size $n$ from the population with two observed variables $X$ and $Y$.

\subsection{Ratio and Product Estimators} \label{subsec:RPE}

It is believed that
%Graunt (1662),
\cite{Grau:1662},
the founder of demography, was the first to propose the use of the ratio estimator to estimate the population of England. %Graunt (1662)
\cite{Grau:1662}
estimated the total population size based on the known annual number of registered births multiplied by the estimated ratio of the population to births. Later in 1802, Laplace \citep{Lapl:1820, Coch:1978}
followed the same approach of
%Graunt (1662)
\cite{Grau:1662}
to estimate the total population of France; however based on a sample of communes (small administrative districts). He used the number of registered births in the preceding year, and estimated the ratio by dividing the total population in the sampled communes to the total number of registered births in the preceding year in same communes. For more details on the origin of the ratio estimator, reader is referred to
%Cochran (1978), Nahar et al. (1993), Sen (1993), and Ahmad et al. (2013b). 
\cite{Coch:1978}, \cite{Naha:Raha:Lask:1993}, \cite{Sen:1993} and \cite{Ahma:Shah:Hani:2013}.

A classical ratio estimator for the population mean ($\mu_Y$) is defined as \citep{Coch:1977};
\begin{equation}
  \hat{\mu}_{Y\!R} = \frac{\bar y}{\bar x} \mu_X \,, \label{eq:ratioE}
\end{equation}
where $\bar y$ and $\bar x$ are the sample means of the variables $Y$ and $X$, respectively, and $\mu_X$ is the known population mean of the auxiliary variable $X$. It is common when the interest is to estimate populations sizes that the auxiliary variable $X$ be the same as the variable $Y$ but in a preceding year; and in this case,
the ratio $\bar y/\bar x$ presents the rate of change over the years. 

The estimator in Equation~\eqref{eq:ratioE} is biased for the population mean ($\mu_Y$) unless the sample size is sufficiently large. The variance of the estimator in Equation~\eqref{eq:ratioE} is always lower than the
variance of the conventional sample mean estimator (i.\,e. $\bar y = \sum_{i=1}^n y_i /n$).
The efficiency of the ratio estimator is at its best when there is a strong positive relationship
between $Y$ and $X$; i.\,e. $\varrho_{Y,X} > \frac12 \frac{C_x}{C_y}$, where $\varrho_{Y,X}$ is the correlation
coefficient between $X$ and $Y$, and $C_x$ and $C_y$ are the coefficients of
variation related to $Y$ and $X$, respectively \citep{Sing:2003}.

The product estimator, on the other hand, is known to be efficient when there is a strong negative linear relationship between $Y$ and $X$; i.\,e. $\varrho_{Y,X} < -\frac12 \frac{C_x}{C_y}$ \citep{Sing:2003}.
The product estimator is defined as
\begin{equation}
	\hat{\mu}_{Y\!P} = \frac{\bar y \bar x}{\mu_X} \,. \label{eq:prodE}
\end{equation}
Several modifications have been made to the ratio and product estimators by integrating different auxiliary variable parameters, such as the coefficient of variation, coefficient of kurtosis, coefficient of skewness, population correlation coefficient, and population deciles. For further details, the reader is referred to
%Walsh (1970), Reddy (1973), Cochran (1977), Gupta (1978), Vos (1980), Sisodia and Dwivedi (1981), Upadhyaya and Singh %(1999), Singh (2003), Kadilar and Cingi (2006), Koyuncu and Kadilar (2009), Yan and Tian (2010), and Subramani and %Kumarapandiyan (2012a, 2012b, 2012c); among many others.
\cite{Wals:1970}, \cite{Redd:1973}, \cite{Coch:1977}, \cite{Gupt:1978}, \cite{Vos:1980}, \cite{Siso:Dwiv:1981},
\cite{Upad:Sing:1999}, \cite{Sing:2003}, \cite{Kadi:Cing:2006}, \cite{Koyu:Kadi:2009},
\cite{Yan:Tian:2010}, \cite{Subr:Kuma:2012a, Subr:Kuma:2012b, Subr:Kuma:2012c}; among many others.

\subsection{Difference and Regression Estimators} \label{subsec:DRE}

The ratio estimator defined in Equation~\eqref{eq:ratioE} can be considered to result from a linear regression equation/line that satisfies three main properties. These are passing through the origin
(zero intercept), slope equaling $\frac{\bar y}{\bar x}$, and passing through the point ($\hat{\mu}_Y, \mu_x$).
It is not always guaranteed, however, that the line relating $Y$ and $X$ passes through the origin. In such cases, the ratio estimator efficiency is degraded. 

To account for this drawback, a more generalized estimator having the form
\begin{equation*}
	\hat{\mu}_Y = \alpha + \beta \mu_X \,,
\end{equation*}
has been proposed to satisfy the linear relationship between $Y$ and $X$ and passes through the point
($\hat{\mu}_Y, \mu_x$) without the necessity to pass through the origin. Substituting with the ordinary least squares estimate for the intercept, we obtain
\begin{equation}
	\hat{\mu}_Y = \bar y + \beta (\mu_X - \bar x) \,. \label{eq:olsE}
\end{equation}
It was found that if $\beta$ is selected to be the population regression coefficient that is equal to
$\frac{\sigma_{XY}}{\sigma_X^2} = \varrho_{Y,X} \frac{\sigma_Y}{\sigma_X}$,
it will provide the minimum variance for the estimator $\hat{\mu}_Y$; where $\sigma_{XY}$
is the population covariance between $Y$ and $X$, and where $\sigma_X$ and $\sigma_Y$
represent the population standard deviations of $Y$ and $X$, respectively.
In this case, the estimator becomes
\begin{equation}
	\hat{\mu}_{Y\!D} = \bar y + \varrho_{Y,X} \frac{\sigma_Y}{\sigma_X} (\mu_X - \bar x) \,, \label{eq:diffE}
\end{equation}
and is referred to as the \textit{difference estimator}.
If the population regression coefficient is unknown and estimated from a sample, then we have
\begin{equation}
  \hat{\mu}_{Y\!Reg} = \bar y + \frac{S_{YX}}{S_X^2} (\mu_X - \bar x) \,, \label{eq:regE}
\end{equation}
where $S_{YX}$ is the sample covariance between $Y$ and $X$, and $S_X^2$ is the sample variance of $X$. 
The estimator in Equation~\eqref{eq:regE} is referred to as the \textit{linear regression estimator}. 
The estimator $\hat{\mu}_{Y\!D}$ defined in Equation~\eqref{eq:diffE} is an unbiased estimator for the population mean, while $\hat{\mu}_{Y\!Reg}$ defined in Equation~\eqref{eq:regE} is biased
%(Cochran (1977); Singh (2003); Singh and Mangat (1996)).
\citep{Coch:1977, Sing:2003, Sing:Mang:1996b}.
The difference and regression estimators are at least as efficient as the ratio and product estimators. 

Similar developments to that for the ratio and product estimators have been done for the difference and regression estimators to increase their efficiency by integrating other auxiliary
variable parameters. There are many auxiliary information based estimators and many of these have been proposed for process monitoring.

\section{Literature Review on Auxiliary Information Based (AIB) SPC/M } \label{sec:RAIB}

In this section we review the literature on using auxiliary information while monitoring process QCs using the various control-charting techniques. We classify the large number of papers on this topic into two sub-sections. The first is devoted to the AIB Phase I control charting techniques, and the second is for the AIB Phase II techniques. Further, the Phase II techniques were divided into the use of univariate charts (Shewhart, EWMA types, and CUSUM types) and the use of multivariate charts.

Generally in the literature, integrating auxiliary information into control charts implies the use of an AIB estimator (similar to those identified in Section~\ref{sec:EuAI}) in place of the conventional estimators (sample mean, variance, standard deviation, \ldots, etc.) to monitor the process parameter(s) of interest. Under the assumptions made, the resulting reduction in variance of the estimation invariably leads to improved control chart performance over the non-AIB counterpart. We can generally conclude that the performance of any AIB method will improve as the correlation between the auxiliary and quality variables increases. This results from the increase of the available information about the quality variable. Similarly, increasing the number of auxiliary variables typically improves performance as the correlation between the auxiliary variables decreases. 

All of the comparisons in the literature are based on the implicit assumption that any process shift, regardless of size, is to be detected quickly. In an increasing number of cases, increasing chart sensitivity can be detrimental. See \cite{Wood:Falt:2019}.

\subsection{Use of Auxiliary Information in Phase I Control Charting Techniques} \label{subsec:AIBphase1}

%Riaz (2008b)
\cite{Riaz:2008b}
was the first to integrate an AIB estimator into a control charting technique. In his study, he assumed a single auxiliary characteristic that is related to the characteristic of interest through a bivariate normal distribution. He used a regression estimator to estimate the process variance instead of the classical sample variance and monitored it using a Shewhart chart. Comparing with a classical Shewhart chart, he reported the higher power the AIB chart
provides especially when $\varrho_{Y,X}$  is high.
%Riaz (2008a)
\cite{Riaz:2008a}
replicated this study on a Phase I Shewhart chart for the process mean. In this case, the regression estimator was used as a replacement for the classical sample mean. In this latter study, he reached the same conclusion regarding the relative performance of the proposed chart to the classical one. 

%Riaz and Does (2009)
\cite{Riaz:Does:2009}
extended
%Riaz’s (2008b)
\citeauthor{Riaz:2008b}'s (\citeyear{Riaz:2008b})
study on the Phase I Shewhart variability chart by using a ratio-type estimator for the process variance instead of the regression estimator, while holding all other assumptions the same. They reported that the use of the ratio estimator improved the chart performance compared to that proposed by
%Riaz (2008b)
\cite{Riaz:2008b}
and its classical counterpart. Recently,
%Abbasi and Adegoke (2020)
\cite{Abbas:Adeg:2020}
considered a wider range of possible AIB estimators for the same chart; which are the regression, ratio, ratio-exponential, power-ratio, ratio-regression, chain ratio cum regression, and chain ratio cum regression exponential estimators. They recommended the ratio exponential estimator as it provided the best overall performance when $\varrho_{Y,X}$ has a moderate or high value. Even when the correlation is low, it provides performance close to that of the classical Shewhart chart. These comparisons are, however, questionable. The ratio-type and regression-type estimators have different requirements for the relationship between the auxiliary variable and variable of interest.
As mentioned in Section~\ref{sec:EuAI},
the ratio-type estimator requires the regression line to pass through the origin, while the regression-type estimators does not. Hence, one cannot arbitrarily use the ratio-type estimators instead of the regression-based one. 

%Riaz et al. (2016)
\cite{Riaz:EtAl:2016}
proposed enhancing the performance of the Shewhart chart designed to monitor the process variability through the dual use of the auxiliary information. A dual use of the auxiliary information implies using the auxiliary variable in ranking and at the estimation stage. They proposed the use of ranked set sampling (RSS) and extreme ranked set sampling (ERSS) schemes, besides the usual simple random sampling. The RSS scheme can be a cost-effective sampling technique that researchers could resort to when the measurement of the units is difficult or expensive. It is usually based on the initial visual ordering of the sample units
%(for more details, readers are referred to McIntyre (1952)).
\citep[for more details, readers are referred to][]{McIn:1952}.
They evaluated the chart assuming the process follows the normal, $t$, or lognormal distribution. In our view, the number of applications of RSS in quality monitoring practice will be quite limited.

%Riaz (2011)
\cite{Riaz:2011} and
%Abbasi and Riaz (2013)
\cite{Abba:Riaz:2013}
extended 
%Riaz’s (2008a)
\citeauthor{Riaz:2008a}'s (\citeyear{Riaz:2008a})
study on the Phase I Shewhart mean chart by using different AIB estimators and increasing the number of the considered auxiliary characteristics.
%Riaz (2011)
\cite{Riaz:2011}
proposed the use of a product-difference type estimator for monitoring the process mean in place of the regression estimator. Due to the noticeably improved performance due to using this mixed estimator, he recommended pooling different styles of auxiliary information use as it improves the chart performance.
%Abbasi and Riaz (2013)
\cite{Abba:Riaz:2013}
investigated the effect of increasing the number of auxiliary variables used. They assumed having two auxiliary characteristics to estimate the process mean using three different AIB estimators. Accordingly, the distribution was changed to a trivariate normal distribution. The estimators used were the
%Abu-Dayyeh et al.’s (2003)
\cite{AbuD:EtAl:2003}
extended two-auxiliary based ratio estimator, the extended two-auxiliary based regression estimator, and
%Kadilar and Cingi’s (2005)
\cite{Kadi:Cing:2005}
two-auxiliary based mixed ratio-regression estimator. The improved performance of their proposed techniques was conditioned on having a moderate to
high value of $\varrho_{Y,X}$, and a weak correlation between the auxiliary characteristics.

%Hussain et al. (2020)
\cite{Huss:EtAl:2020c}
designed an AIB median control chart for the process location parameter in Phase I. The importance of a median chart over its mean counterpart is that the former is more efficient when process outliers are expected, and/or the process distribution is significantly skewed. Using one auxiliary characteristic,
%Hussain et al. (2020)
\cite{Huss:EtAl:2020c}
compared between the classical sample median, classical ratio estimator, and other ratio-type estimators based on a known coefficient of variation and coefficient of kurtosis of the auxiliary variable. They reported that the ratio-type estimators based on the coefficient of variation and kurtosis were preferable.

\subsection{Use of Auxiliary Information in Phase II Control Charting Techniques} \label{subsec:AIBphase2}

\subsubsection{Univariate Control Charts}

\paragraph{Shewhart Chart and its Extended Versions} 

%Riaz et al. (2013)
\cite{Riaz:EtAl:2013}
studied the AIB-Shewhart chart for monitoring the process mean in both Phase I and Phase II under the assumption that the process follows either a normal, gamma, or $t$- distribution. In addition to the classical sample mean, they assessed six different AIB estimators; which are the product, ratio, regression, product-difference used in
%Riaz (2011),
\cite{Riaz:2011}
and two ratio estimators based on the coefficient of variation and kurtosis of the auxiliary characteristic. Their comparison revealed the better performance of the regression estimator.
%Riaz (2015a)
\cite{Riaz:2015a}
extended this study under the normal and $t$ Phase II distributions; while comparing between other AIB estimators depending on either one or two auxiliary variables. They further added the Hotelling’s $T^2$ chart to their comparison, despite its different intended purpose. They concluded that the best performance is achieved if two auxiliary characteristics are used with the power-ratio estimator
%(proposed by Abu-Dayyeh et al. (2003))
\citep[proposed by][]{AbuD:EtAl:2003}
or the modified ratio-type estimator; both being based on the coefficient of variation of the auxiliary characteristic. 

%Lee et al. (2015)
\cite{Lee:EtAl:2015}
proposed the use of repetitive sampling technique on the AIB-Shewhart chart with the product-difference type estimator proposed by
%Riaz (2011).
\cite{Riaz:2011}.
The repetitive sampling technique allows for taking additional samples at a given time point when the results of the drawn sample are indecisive.
%Lee et al. (2015)
\cite{Lee:EtAl:2015}
reported better performance for their chart compared to that of
%Riaz (2011),
\cite{Riaz:2011},
especially when $\varrho_{Y,X}$ is low.
%Abbasi and Riaz (2016)
\cite{Abba:Riaz:2016}
suggested making dual use of the auxiliary information for an AIB-Shewhart chart using the regression estimator. They proposed the use of RSS, median RSS (MRSS), and ERSS schemes.
%Mehmood et al. (2017)
\cite{Mehm:EtAl:2017}
extended the consideration of dual use of auxiliary information on location charts using RSS, ERSS, double RSS (DRSS), and double ERSS (DERSS) sampling schemes for known process distribution and unknown skewed process distribution.

%Haq and Khoo (2016)
\cite{Haq:Khoo:2016}
developed an AIB synthetic chart for monitoring the process mean. A synthetic chart, introduced by
%Wu and Spedding (2000),
\cite{Wu:Sped:2000a},
is an integration of the Shewhart and conforming run length (CRL) charts, basically an application of a runs rule. As did
%Knoth (2016),
\cite{Knot:2016a},
we strongly recommend against the use of the synthetic control charts, however, because of their generally poor performance when fairly compared to other competing charts using steady-state run length properties. Using the difference estimator, the performance of the chart proposed by
%Haq and Khoo (2016)
\cite{Haq:Khoo:2016}
was found better than the classical synthetic chart for all correlation ($\varrho_{Y,X}$) levels and the classical EWMA chart for
$\varrho_{Y,X} \ge 075$.
%Saha et al. (2019)
\cite{Saha:EtAl:2019}
evaluated a variable sample size and sample interval (VSSI) chart for monitoring the process mean while integrating some auxiliary information. The VSSI chart introduced by
%Prabhu et al. (1994)
\cite{Prab:EtAl:1994}
depends on altering the sample size and sampling interval based on the process quality status reported from the prior sample. Based on the difference estimator, the AIB-VSSI chart showed better performance than its non-AIB counterpart, the conventional EWMA chart, the AIB-EWMA chart, and the AIB-synthetic chart. The flexibility of the VSSI approach provides better statistical performance, but we see little evidence of its use in practice due to the added complexity and the resulting varying workload. 

%Haq and Khoo (2018)
\cite{Haq:Khoo:2018}
evaluated an AIB double sampling (DS) Shewhart chart for the process mean using the difference estimator. The DS chart, introduced by
%Daudin (1992),
\cite{Daud:1992},
allows for drawing a second sample from the process if the first sample results were indecisive. The process status is determined based on the pooled results from both samples. The DS can be considered as a special case from the repetitive sampling technique with a limitation of two samples at most being drawn at a given time point. They reported that their proposed AIB-DS chart is at least as efficient as the classical DS chart, and outperforms the conventional EWMA chart only for high values of $\varrho_{Y,X}$ ($\varrho_{Y,X} \ge 0.75$) and specific shift sizes.
%Haq and Khoo (2019a)
\cite{Haq:Khoo:2019a}
integrated a CRL chart with the AIB-DS chart proposed by
%Haq and Khoo (2018)
\cite{Haq:Khoo:2018}
to develop an AIB synthetic DS control chart. Using the difference estimator as well, they showed that an optimally designed AIB synthetic DS chart outperforms the AIB-Shewhart, optimal AIB-synthetic, and DS charts. Further,
%Umar et al. (2020)
\cite{Umar:EtAl:2020}
proposed adding the variable sampling interval (VSI) feature to the AIB-DS chart for the process mean. The VSI chart, proposed by
%Reynolds et al. (1988),
\cite{Reyn:EtAl:1988},
is a special case of the VSSI chart in which only the sampling interval is adaptable. Using the difference estimator, they reported that their proposed chart outperforms the AIB-VSSI chart proposed by
%Saha et al. (2019)
\cite{Saha:EtAl:2019}
and the AIB-DS chart proposed by
%Haq and Khoo (2018).
\cite{Haq:Khoo:2018}.
However, it has a lower performance for small shift sizes when compared to an AIB-EWMA and AIB-Run Sum (RS) charts. The AIB-RS chart was proposed by
%Ng et al. (2018),
\cite{Ng:EtAl:2020a}
who reported its superior performance over that of the AIB-Shewhart, AIB-synthetic, and AIB-EWMA charts
under the condition that $\varrho_{Y,X}$ is considerably high.

The Phase II Shewhart median chart was also evaluated when integrated with auxiliary information. Under the normality assumption,
%Ahmad et al. (2014b)
\cite{Ahma:EtAl:2014b}
evaluated the use of one and two auxiliary characteristics to estimate the process median; depending on some estimators based on the coefficient of variation and kurtosis of the auxiliary characteristics. They recommended the use of two auxiliary variables over the use of one.
%Ahmad et al. (2014c)
\cite{Ahma:EtAl:2014c}
evaluated a median-based Shewhart chart under the assumption of normal, $t$, and gamma-distributed processes using
a DS scheme integrated with some AIB estimators.

%Ahmad et al. (2019)
\cite{Ahma:EtAl:2019}
considered the use of auxiliary information in univariate and bivariate autoregressive processes of the first order, i.e., AR(1) processes. They applied different proposed AIB estimators, and compared them with the classical estimator. The AIB estimators showed an improved performance with a recommendation for the use of the modified Hartley and Ross estimator.

Monitoring the process variability also has been considered in the AIB monitoring literature.
%Ahmad et al. (2013a)
\cite{Ahma:EtAl:2013}
studied the Shewhart variability chart based on one or two auxiliary characteristics under simple random sampling and DS techniques. Both normal and gamma distributions were considered under uncontaminated and contaminated conditions. They recommended the use of two auxiliary characteristics.
%Riaz et al. (2014)
\cite{Riaz:EtAl:2014}
evaluated several AIB estimators for the process dispersion; including the ratio, power ratio, ratio exponential, ratio regression, power ratio regression, and ratio exponential regression methods. They recommended the use of the regression and power ratio estimators.
%Ahmad et al. (2014a)
\cite{Ahma:EtAl:2014a}
evaluated this same chart based on two auxiliary characteristics, while assuming normal and $t$ distributed processes. They compared the usual variance estimator, ratio and regression estimators with a two auxiliary-based regression estimator, chain ratio type estimator, and chain ratio cum regression type estimator. Generally, they recommended the use of two auxiliary characteristics and a chain ratio type estimator under a normal distribution and with highly correlated variables under a $t$ distribution. Other dispersion measures were also proposed to monitor the process variability in conjunction with other auxiliary parameters; such as the CV and interquartile range (IQR). See, for example,
%Ahmad et al. (2012), Riaz (2015b), and Abbasi (2020). 
\cite{Ahma:EtAl:2012}, \cite{Riaz:2015b} and \cite{Abbasi:2020}.

\paragraph{AIB EWMA Chart and its Extended Versions}

%Abbas et al. (2014)
\cite{Abba:Riaz:Does:2014}
were the first to propose integrating auxiliary information into an EWMA charting technique. They used the difference estimator (termed in their article regression estimator) using a single auxiliary variable to estimate the mean of the process. Their proposed technique showed better performance than the classical EWMA chart, especially when the value of $\varrho_{Y,X}$ is high. It also showed better performance compared to a classical CUSUM and multivariate EWMA (MEWMA) charts over almost all levels of $\varrho_{Y,X}$.
%Noor-ul-Amin et al. (2018)
\cite{Noor:Khan:Asla:2018}
consolidated two parametric ratio estimators to monitor the process mean using an EWMA chart. Their estimator was a function of the coefficient of kurtosis and quartile deviation of the auxiliary characteristic. The quartile deviation is defined as half the IQR. Since their suggested estimator is of the ratio type, it was then conditioned on having a positive value for $\varrho_{Y,X}$.
In comparison with the chart proposed by
%Abbas et al. (2014),
\cite{Abba:Riaz:Does:2014},
their chart provided improved performance.
%Zichuan et al. (2020)
\cite{Zich:EtAl:2020}
evaluated the AIB-EWMA chart using two auxiliary characteristics; assuming either the presence or absence of collinearity between them. Based on the extended regression estimator of
%Kadilar and Cingi (2005),
\cite{Kadi:Cing:2005},
their proposed technique showed higher detection ability when there was no collinearity between the auxiliary variables.
%Ng et al. (2021)
\cite{Ng:EtAl:2021}
evaluated the AIB-EWMA chart proposed by
%Abbas et al. (2014)
\cite{Abba:Riaz:Does:2014}
along with the AIB-Shewhart chart proposed by
%Riaz (2008a)
\cite{Riaz:2008a}
and the AIB-synthetic chart proposed by
%Haq and Khoo (2016)
\cite{Haq:Khoo:2016}
in terms of the economic and economic-statistical design performance.
%Ng et al. (2021)
\cite{Ng:EtAl:2021}
are considered the first to tackle these types of performance evaluations with AIB charts. In general, we find that the economic-statistical design approach has substantial advantages over the economic design approach. 

Further research was conducted on the use of different sampling schemes to enhance the chart performance. For example,
%Saghir et al. (2019)
\cite{Sagh:EtAl:2019}
followed the same repetitive sampling approach of
%Lee et al. (2015)
\cite{Lee:EtAl:2015}
for an AIB-EWMA chart designed with a product-difference estimator.
%Adegoke et al. (2017)
\cite{Adeg:EtAl:2017}
proposed the use RSS and MRSS for an AIB-EWMA chart based on a product estimator.
%Adegoke et al. (2017)
\cite{Adeg:EtAl:2017}
study was restricted on having a negative value for $\varrho_{Y,X}$.
%Adegoke et al. (2019)
\cite{Adeg:EtAl:2019c}
extended the study by using the regression estimator instead. 

%Arshad et al. (2017)
\cite{Arsh:EtAl:2017}
extended the study on the AIB-EWMA chart for monitoring the process mean by adding run rules in an attempt to increase its sensitivity. However, this is usually not recommended and we do not support using runs rules with either the EWMA or CUSUM charts. They proposed nine different signaling rules based on using the difference estimator (termed in their article regression estimator) and single auxiliary characteristic. They compared their methods with some AIB- and non-AIB EWMA and CUSUM charts.
%Ng et al. (2020)
\cite{Ng:EtAl:2020b}
added the VSI feature to the AIB-EWMA chart proposed by
%Abbas et al. (2014);
\cite{Abba:Riaz:Does:2014};
namely the VSI AIB-EWMA chart. Using the difference estimator, the proposed chart provided improved performance, and its sensitivity increases with increases in the value of $\varrho_{Y,X}$. Recently,
%Umar et al. (2021)
\cite{Umar:EtAl:2021}
optimally designed the chart proposed by
%Ng et al. (2020)
\cite{Ng:EtAl:2020b}
based on the average performance over a range of mean shifts.

%Haq (2018)
\cite{Haq:2018}
proposed integrating the auxiliary information technique with an adaptive EWMA chart proposed by
%Haq et al. (2018).
\cite{Haq:Gulz:Khoo:2018}.
With this adaptive EWMA chart one first estimates the unknown mean shift using an EWMA statistic, and based on this estimate the smoothing parameter of the EWMA chart is determined.
%Haq (2018)
\cite{Haq:2018}
proposed the use of the difference estimator to estimate the shift. Based on his results, the proposed chart outperforms its non-AIB counterpart, in addition to the classical EWMA, CUSUM, synthetic EWMA, and synthetic CUSUM charts. Recently,
%Haq et al. (2021a)
\cite{Haq:Akht:Khoo:2021}
proposed adding a VSI feature to the AIB-adaptive EWMA chart and found it more sensitive than its fixed sampling interval (FSI) counterpart.
%Anwar et al. (2020a)
\cite{Anwa:EtAl:2020a}
evaluated the modified–EWMA (MxEWMA) chart introduced by
%Khan et al. (2017)
\cite{Khan:Asla:Jun:2017}
when the difference estimator is used (termed in their article regression estimator). The MxEWMA chart adds an extra component to the chart statistic based on the difference between the current and previous sample statistic.

Extended versions of the EWMA chart for the process mean were also considered while using AIB estimators. For example,
%Haq et al. (2019)
\cite{Haq:Abid:Khoo:2019}
evaluated an EWMA-$t$ chart using the difference estimator (termed in their article regression estimator). The EWMA-$t$ chart, introduced by
%Zhang et al. (2009),
\cite{Zhan:Chen:Cast:2009},
is used to protect against the instability of the process variance when the mean is stable. The $t$-statistic refers to the test-statistic of the one-sample Student’s $t$-test. When  increases, the AIB-EWMA-$t$ outperforms its non-AIB version. Yet, an AIB-EWMA chart is significantly more sensitive than an AIB-EWMA-$t$ chart as long as the variance is stable. Recently,
%Haq et al. (2021b)
\cite{Haq:EtAl:2021b}
proposed imposing a FSI and VSI features to a double AIB-EWMA-$t$ chart. Their results showed an improved performance comparing to the AIB-EWMA-$t$ and DEWMA-$t$ charts.

%Chen and Lu (2020)
\cite{Chen:Lu:2020}
proposed a generally weighted moving average (GWMA)-$t$ and an AIB-GWMA-$t$ charts for monitoring the process mean using the regression estimator. They concluded that the AIB-GWMA-$t$ chart performs uniformly and substantially better than the EWMA-$t$ and GWMA-$t$ charts, and if it is designed with large parameter values, it outperforms the AIB-EWMA-$t$ chart within a certain range of the correlation ($\varrho_{Y,X}$).
%Lu et al. (2021)
\cite{Lu:Chen:Yang:2021}
evaluated an AIB-MaxGWMA chart that simultaneously monitors the process mean and variance. There have been recent recommendations against the use of the GWMA charts
%(Knoth et al. (2021b))
\citep{Knot:EtAl:2021b}
which we support.
%Noor-ul-Amin et al. (2019a)
\cite{Noor:Khan:Sana:2019a}
proposed an AIB hybrid exponential weighted moving average (AIB-HEWMA) chart by using two-parametric ratio estimator
%(same as that used in Noor-ul-Amin et al. (2018))
\cite[same as that used in][]{Noor:Khan:Asla:2018}
for the process mean.
%Haq (2013)
\cite{Haq:2013}
proposed the HEWMA chart, which is equivalent to the double EWMA (DEWMA) chart introduced by
%Shamma et al. (1991) and Shamma and Shamma (1992).
\cite{Sham:Amin:Sham:1991} and \cite{Sham:Sham:1992}.
Note that \cite{Zhan:Chen:2005} also proposed an equivalent chart.
%Mahmoud and Woodall (2010)
\cite{Mahm:Wood:2010}
recommended against the use of the DEWMA chart because, for one reason, it gives greater weight to past data values than to current values. Such weighting patterns are unreasonable and lead to poor steady-state run length performance.
%Raza et al. (2019)
\cite{Raza:EtAl:2019}
evaluated the HEWMA chart using exponential type estimator of the mean based on two auxiliary characteristics.

%Adegoke et al. (2019)
\cite{Adeg:EtAl:2019a}
integrated the auxiliary information to the homogenously weighted moving average (AHWMA) chart.
%Abbas (2018)
\cite{Abba:2018}
introduced the HWMA chart, where the recent sample observation is assigned to a certain weight, while the rest of the weights are distributed equally on the historical observations. Recently,
%Knoth et al. (2021a)
\cite{Knot:EtAl:2021a}
criticized the use of the HWMA chart and showed that it had poor performance in detecting delayed shifts in the process.
%Anwar et al. (2021a)
\cite{Anwa:EtAl:2021a}
evaluated an AIB double version of the HWMA chart (DHWMA) using the difference estimator.
%Amir (2021b)
\cite{Amir:EtAl:2021b}
proposed the use of the difference estimator when evaluating the moving average (MA) chart, while
%Amir et al. (2021a)
\cite{Amir:EtAl:2021a}
proposed the use of the same estimator when evaluating the double moving average (DMA) chart for the process mean. We recommend against using the double moving average method because it also gives more weight to past data values than to the current values, an undesirable property for process monitoring. 

%Hussain et al. (2018)
\cite{Huss:EtAl:2018}
considered a median-based EWMA chart using one and two auxiliary characteristics. Based on their results, the use of two auxiliary characteristics has an overall dominant performance, and unsurprisingly the median based chart was mainly preferable when contamination exists.
%Haq (2020)
\cite{Haq:2020}
had the first research article on exploring the AIB-nonparametric control charts with an application based on the EWMA sign chart.

Monitoring the process variability with AIB EWMA charts has also been studied.
%Haq (2017b)
\cite{Haq:2017b}
designed an AIB-EWMA chart for dispersion using a difference estimator (termed in his article regression estimator).
%Abbasi et al. (2020)
\cite{Abbasi:EtAl:2020b}
assessed the use of different AIB estimators on the performance of the EWMA chart for monitoring the process variability. They proposed the use of ratio, regression, power ratio, ratio exponential, ratio regression, power ratio regression, and ratio exponential regression estimators. Based on the chart performance, they recommended the use of the regression, power ratio, and ratio exponential estimators.
%Hussain et al. (2019b)
\cite{Huss:EtAl:2019b}
assessed an AIB-EWMA chart for dispersion based on the IQR.
%Nuriman et al. (2021)
\cite{Nuri:Mash:Ahsa:2021}
evaluated a generalized weighted moving average chart for the coefficient of variation (AIB-GWMCV) using the difference estimator.

Also, the AIB-EWMA chart was extended to monitor simultaneously the mean and the dispersion of a process. For example,
%Haq (2017a)
\cite{Haq:2017a}
evaluated an AIB-Maximum EWMA (AIB-MaxEWMA) chart that can simultaneously monitor both parameters.
%Chen and Cheng (1998)
\cite{Chen:Chen:1998}
introduced the Max Chart, and
%Chen et al. (2001)
\cite{ChenG:ChenS:Xie:2001}
extended it to the EWMA chart. Since the MaxEWMA chart combines the EWMA mean and the EWMA dispersion charts into a single chart,
%Haq (2017a)
\cite{Haq:2017a}
used the difference estimator for each of the process mean and variance while depending on a single auxiliary characteristic. The AIB chart was reported to be uniformly better than its non-AIB counterpart.
%Haq and Akhtar (2021)
\cite{Haq:Akht:2021}
integrated the AIB and VSI features to the MaxEWMA and MaxDEWMA charts.
%Noor-ul-Amin (2021)
\cite{Noor:EtAl:2021}
investigated the effect of having measurement errors on the performance of the AIB-MaxEWMA chart.
%Noor-ul-Amin (2019b)
\cite{Noor:Tari:Hani:2019b}
proposed similar MaxEWMA and AIB-MaxEWMA charts for simultaneously monitoring the process mean and coefficient of variation.
%Javaid et al. (2020)
\cite{Java:Noor:Hani:2020}
proposed an AIB maximum HEWMA (MaxHEWMA) chart for joint monitoring of process mean and variance. Using the difference estimator for the mean and the variance, the chart becomes sensitive to small shifts in process mean and variance, even more so than the AIB-maxEWMA chart.
%Javaid et al. (2021)
\cite{Java:Noor:Hani:2021}
evaluated this chart in the presence of measurement errors.

\paragraph{AIB CUSUM Charts and their Extended Versions}

%Sanusi et al. (2017)
\cite{Sanu:EtAl:2017}
developed an AIB combined Shewhart-CUSUM chart using a single auxiliary characteristic under the normality assumption. They proposed using the regression, ratio, Singh and Tailor, power ratio, and the
%Kadilar and Cingi (2005)
\cite{Kadi:Cing:2005}
estimators for estimating the location parameter. They recommended using the power-ratio type estimator if the value of $\varrho_{Y,X}$ is low,
or the ratio estimator/Singh and Tailor estimator if the value of $\varrho_{Y,X}$ is high.
%Haq (2017c)
\cite{Haq:2017c}
assessed the synthetic EWMA and synthetic CUSUM for the process mean under the normality assumption using the difference estimator. The proposed charts were shown to have better detection ability to process mean changes than their non-AIB-based counterparts and the conventional EWMA and CUSUM charts.

%Sanusi et al. (2018)
\cite{Sanu:Abba:Riaz:2018}
integrated the auxiliary information in a conventional CUSUM chart for monitoring the process mean. Under the normality assumption, they evaluated nine different AIB estimators. The regression, ratio, Singh and Taylor, and power ratio estimators provided a quite reasonable performance. Classical EWMA and CUSUM charts were compared to the proposed chart and results revealed the better performance of the latter.
%Abbasi and Haq (2019b)
\cite{Abba:Haq:2019b}
proposed an AIB optimal CUSUM (AIB-OCUSUM) and an AIB adaptive CUSUM (AIB-ACUSUM) chart using different AIB estimators for the process mean. The OCUSUM chart is the chart optimized to detect efficiently a specific shift size. With the ACUSUM one first estimates the mean shift size using an EWMA statistic and accordingly updates the CUSUM chart reference
parameter. The ACUSUM chart is known for its good detection power over a range of shift sizes.
%Abbasi and Haq (2019a)
\cite{Abba:Haq:2019a}
developed two AIB versions from the improved ACUSUM (IACUSUM) introduced by
%Abbasi and Haq (2020)
\cite{Abba:Haq:2020}
for the process mean. The first version (AIB-IACUSUM) uses an unbiased EWMA statistic to estimate the mean shift and then updates the reference parameter of the CUSUM chart. The second version (AIB-IACCUSUM) replaces the CUSUM chart in the IACUSUM chart with the Crosier’s CUSUM chart
\citep{Cros:1986}. Additionally,
%Haq et al. (2021a)
\cite{Haq:Akht:Khoo:2021}
developed an AIB adaptive Crosier CUSUM (AIB-ACCUSUM) for the process mean.

Based on the difference estimator,
%Haq and Abidin (2019)
\cite{Haq:Abid:2019}
evaluated an AIB-CCUSUM-$t$ chart for the process mean, along with incorporating the fast initial response (FIR) feature to increase its sensitivity in detecting initial process changes.
%Haq and Bibi (2021)
\cite{Haq:Bibi:2021}
used the difference estimator as well to develop AIB dual CUSUM (AIB-DCUSUM) and AIB dual Crosier CUSUM (AIB-DCCUSUM) charts with and without the FIR feature. A dual CUSUM chart uses two separate CUSUM charts to protect against different shift sizes. Their results showed the better performance of the proposed charts over the DCUSUM, DCCUSUM and AIB-CUSUM charts.

%Riaz et al. (2019)
\cite{Riaz_EtAl:2019}
evaluated the effect of having measurement errors on an AIB mixed EWMA-CUSUM chart. The mixed EWMA-CUSUM, proposed by
%Riaz et al. (2013),
\cite{Abba:Riaz:Does:2013a},
uses an EWMA statistic as the input in the CUSUM chart statistic. Generally, we find the justification for these mixed charts to be very weak in that comparisons were made to unnecessarily weak competitors. The AIB mixed EWMA-CUSUM bases the EWMA statistic on the difference estimator (termed in their article regression estimator).
%Anwar et al. (2020b)
\cite{Anwa:EtAl:2020b}
proposed the use of the difference estimator (also termed in their article as regression estimator) with two mixed-types of control charts; the mixed EWMA-CUSUM and mixed CUSUM-EWMA charts. The mixed CUSUM-EWMA chart, proposed by
%Zaman et al. (2015)
\cite{Zama:EtAl:2015}, uses a CUSUM statistic as an input in the EWMA chart statistic. An AIB version form of a mixed CUSUM-EWMA chart implies calculating the CUSUM statistic based on an AIB estimator. Following that,
%Anwar et al. (2021b)
\cite{Anwa:EtAl:2021b}
integrated the auxiliary information using the difference estimator to the combined mixed EWMA-CUSUM chart proposed by
%Zaman et al. (2016)
\cite{Zama:Riaz:Lee:2016}
to simultaneously monitor the location and dispersion process parameters. The chart depends on using two EWMA statistics, one for the mean and another for dispersion as inputs into the classical CUSUM chart.
%Haq and Khoo (2021)
\cite{Haq:Khoo:2021}
developed a multiple AIB (MAIB) estimator that was shown to be a unique uniformly minimum variance unbiased estimator (UMVUE), and used it to evaluate four different memory-type control charts with FSI and VSI features. The four charts are CUSUM, EWMA, AEWMA, and ACUSUM. They recommended the use of their proposed MAIB-charts over AIB charts and non-AIB charts in monitoring the process mean and that the proposed charts’ performance improves with the increase of the number of auxiliary characteristics.

%Hussain et al. (2019a)
\cite{Huss:EtAl:2019a}
proposed the use of auxiliary information for a CUSUM median chart. Symmetric and asymmetric process distributions were considered, while assuming either the presence or absence of the outliers. One and two available auxiliary characteristics were also assumed. They recommended the use of two auxiliary characteristics, and the use of the median chart under contaminated environments and/or skewed distributions.

\subsubsection{Multivariate Control Charts}

%Haq and Khoo (2019b)
\cite{Haq:Khoo:2019b}
introduced the use of auxiliary information in multivariate EWMA (MEWMA), double MEWMA (DMEWMA), and multivariate CUSUM (MCUSUM) charts for monitoring the process mean. Again, we do not recommend the use of the DMEWMA chart for reasons similar to those for its univariate counterpart. They assumed that $q$ normally distributed variables are available on the process by which $p$ of them represent the characteristics of interest, and $q-p$ represent the auxiliary characteristics. The estimator they used is more of a generalized version of the difference estimator. Their proposed charts were compared with their non-AIB components, and showed a uniformly and substantially better performance for different shifts in the process mean vector.

%Haq et al. (2020)
\cite{Haq:Munir:Shah:2020}
used an AIB mean estimator with a dual multivariate CUSUM (DMCUSUM) chart and mixed DMCUSUM (MDMCUSUM) chart assuming the availability of multiple auxiliary characteristics. The DMCUSUM chart combines two similar-type MCUSUM charts with two different reference values and decision intervals to detect efficiently a range of mean shifts. The MDMCUSUM combines two different-type of MCUSUM charts into a single chart; one for the small-to-moderate shifts and the other for moderate-to-large shifts. The two AIB-charts were evaluated with and without FIR features.

%Haq et al. (2021c)
\cite{Haq:EtAl:2021c}
developed an AIB adaptive multivariate EWMA (AIB-AME) and AIB weighted adaptive multivariate CUSUM (AIB-WAMC). They evaluated the proposed charts with and without the VSI feature. Generally, they reported that the AIB-charts are more efficient than their non-AIB counterparts, and adding the VSI feature increases the chart’s sensitivity.

\section{Issues in Applying AIB Estimation Techniques in Quality Monitoring} \label{sec:AIBissues}

Across the studies conducted on the AIB control charts, we can see that researchers impose two main assumptions. The first assumption is that the distributional properties of the auxiliary characteristic are known and unchangeable during the monitoring time. The second assumption is that the AIB charts are designed to monitor processes without a cascade property; and hence any change in the mean of any characteristic would not affect the the mean of others.

The first assumption is unrealistic in many quality control and monitoring applications. Variables in manufacturing processes tend to be subject to assignable causes of variation that change in some way the distribution of measured variables. In none of the work on AIB charts do authors mention the need for a reliable control plan or system for controlling the auxiliary characteristic parameters to guarantee long-term stability. In addition, none of the researchers mention the need for monitoring the auxiliary variables.

As for the second assumption, it is assumed that a change in the distribution of $Y$ will not cause a change in the distribution of $X$ and vice versa. The assumption actually follows from the first in that the distribution of the auxiliary variable cannot change at all. 

In this section, we show two possible cases where a change in the auxiliary characteristic ($X$) could drastically affect the monitoring of the characteristic of interest ($Y$). This could be considered to be a study of the robustness of the AIB methods to a violation of the required assumptions. We consider two control charts, the AIB-Shewhart and AIB-EWMA charts. Both charts are designed based on the difference estimator defined in Equation~\eqref{eq:diffE}. It is assumed that $Y$ and $X$ follow
$\mathcal{N}_2(\mu_{Y0}, \mu_{X0}, \sigma_Y, \sigma_X, \varrho_{Y,X})$, where $\mathcal{N}_2(\cdot)$ refers to a
bivariate normal distribution. The AIB-Shewhart chart statistic at time $i$ is defined as;
\begin{equation}
  Z_i = \bar y_i + \varrho_{Y,X} \frac{\sigma_Y}{\sigma_x} (\mu_x - \bar x_i) \quad,\; i = 1, 2, 3, \ldots \label{eq:AIBshew}
\end{equation}
The chart signals when $Z_i$ exceeds the control limits defined as
\begin{equation*}
  \mu_{Y0} \pm L_s \sqrt{1-\varrho_{Y,X}^2} \, \frac{\sigma_X}{\sqrt{n}} \;,
\end{equation*}
where $L_s$ is chosen to satisfy a specified in-control performance. On the other hand, the AIB-EWMA statistic at time $i$ is defined as
\begin{equation}
  w_i = \lambda Z_i + (1-\lambda) w_{i-1} \quad,\; i = 1, 2, 3, \ldots \qquad (w_0 = \mu_{Y0}) \,, \label{eq:AIBewma}
\end{equation}
where $Z_i$ is defined in Equation~\eqref{eq:AIBshew}, and $\lambda$ is a smoothing parameter such that $0 <\lambda\le 1$.
The chart signals when $w_i$ exceeds the control limits defined as;
\begin{equation*}
	\mu_{Y0} \pm L_e \sqrt{\frac{\lambda}{2-\lambda} (1-\varrho_{Y,X}^2)} \, \frac{\sigma_X}{\sqrt{n}} \;,
\end{equation*}
where $L_e$ is chosen to satisfy a specified in-control performance.
In our study both charts are designed to achieve an in-control average run length (ARL) of 200.
All results presented next were obtained using 50,000 simulation runs, and without loss of generality results are valid for any in-control parameter values for $Y$ and $X$ and any sample size $n$. 

\noindent\textbf{Case I:} False alarm rate

In this case, we show that a change in the auxiliary characteristic ($X$) could yield a misleading mean estimate for the characteristic of interest $Y$; in which case the AIB chart would signal a false alarm while, in fact, $Y$ is in-control. Table~\ref{tab:01} presents the ARL values corresponding to the case when the mean of $Y$ is in-control ($\mu_{Y0}$), while the mean of $X$ shifted from $\mu_{X0}$ to
$\mu_{X1} = \mu_{X0} + \delta_X \sigma_X / \sqrt{n}$\,.
That is, the Phase II observations are generated from
$\mathcal{N}_2(\mu_{Y0}, \mu_{X1}, \sigma_Y, \sigma_X, \varrho_{Y,X})$, 
while the chart statistics of both charts are calculated using \eqref{eq:AIBshew} with $\mu_x = \mu_{X0}$.
As shown, the in-control ARL values become significantly lower than the value of 200 (i.e. the nominal in-control ARL value when $Y$ does not experience a change); especially as $\varrho_{Y,X}$ increases.
This indicates a substantial increase in the rate of false alarms for relatively small changes in the mean of the auxiliary variable. As can be noticed, the only case when the ARL is close to its nominal value (200) is when the correlation between $Y$ and $X$ is very weak
(e.\,g., $\varrho_{Y,X}=0.05$). These results hold for both the Shewhart chart and the EWMA chart with different smoothing parameters ($\lambda$).
\begin{table}[hbt]
\centering
\caption{ARL values of the Shewhart and EWMA control charts when the mean of auxiliary characteristic ($X$) is shifted while the mean of the characteristic of interest is in-control at different levels of correlation (Case I)}\label{tab:01}

\vspace*{1ex}
\begin{tabular}{cc|c|cccc} \toprule
  \raisebox{-2ex}[0mm][0mm]{$\varrho_{Y,X}$} & \raisebox{-2ex}[0mm][0mm]{$\delta_X$}
  & Shewhart & \multicolumn{4}{c}{EWMA with ($\lambda, L_e$)} \\
  & & ($L_s=2.807$) & ($0.05, 2.216$) & ($0.10, 2.454$) & ($0.20, 2.636$) & ($0.50, 2.777$) \\ \midrule
       & 0.25 & 200.4 & 199.5 & 198.4 & 200.1 & 199.9 \\
  0.05 & 0.50 & 198.2 & 193.9 & 194.6 & 196.7 & 198.6 \\
       & 1.00 & 198.1 & 177.8 & 183.2 & 188.8 & 195.0 \\ \midrule
       & 0.25 & 197.5 & 166.9 & 173.0 & 182.5 & 190.9 \\
  0.25 & 0.50 & 186.3 & 111.6 & 124.6 & 142.7 & 168.6 \\
       & 1.00 & 153.4 &  52.4 &  60.3 &  74.9 & 110.7 \\ \midrule
       & 0.25 & 184.1 & 101.8 & 113.9 & 131.9 & 161.6 \\
  0.50 & 0.50 & 145.3 &  45.4 &  51.4 &  65.0 & 100.4 \\
       & 1.00 &  75.6 &  18.3 &  18.3 &  21.0 &  35.9 \\ \midrule 
       & 0.25 & 145.8 &  46.6 &  52.9 &  66.7 & 101.8 \\
  0.75 & 0.50 &  77.6 &  18.6 &  18.9 &  21.8 &  36.9 \\
       & 1.00 &  21.3 &   8.1 &   7.3 &   6.9 &   8.9 \\ \bottomrule        
\end{tabular}
\end{table}

\noindent\textbf{Case II:} Masking process changes

In this case, we show with a simple derivation that when both the auxiliary characteristic $X$ and the characteristic of interest $Y$ change simultaneously, a certain magnitude of change in $X$ can entirely mask the change occurring in $Y$. Consider for example, the extensively used estimator defined in Equation~\eqref{eq:olsE}. We consider a change in the mean of $X$  of size $\Delta_x$
and a change in the mean of $Y$ of size $\Delta_y$. These changes would be reflected in the sample means.
Hence, this implies that
\begin{align*}
  \hat{\mu}_Y^* & = (\bar y + \Delta_y) + \beta \big(\mu_X -(\bar x + \Delta_x)\big) \\
  & = \bar y + \beta (\mu_X - \bar x) + ( \Delta_y - \beta \Delta_x ) \quad.
\end{align*}
If the quantity $\Delta_y - \beta \Delta_x = 0$, then the estimator $\hat{\mu}_Y^*$ will be reduced to $\hat{\mu}_Y$;
i.\,e. the estimator calculated under the in-control parameters of $Y$ and $X$. In such a case, the change that occurred in $Y$ will be masked by the change occurring in $X$. This will be satisfied when $\Delta_x = \Delta_y/\beta$.
Hence, if both parameters of $Y$ and $X$ changed, such that the change in $X$ was equivalent to the change in $Y$ multiplied by the reciprocal of the slope, then practitioners would not detect changes in $Y$.

\begin{figure}[hbt]
\centering
\vspace*{-6ex}
\includegraphics[width=0.57\textwidth]{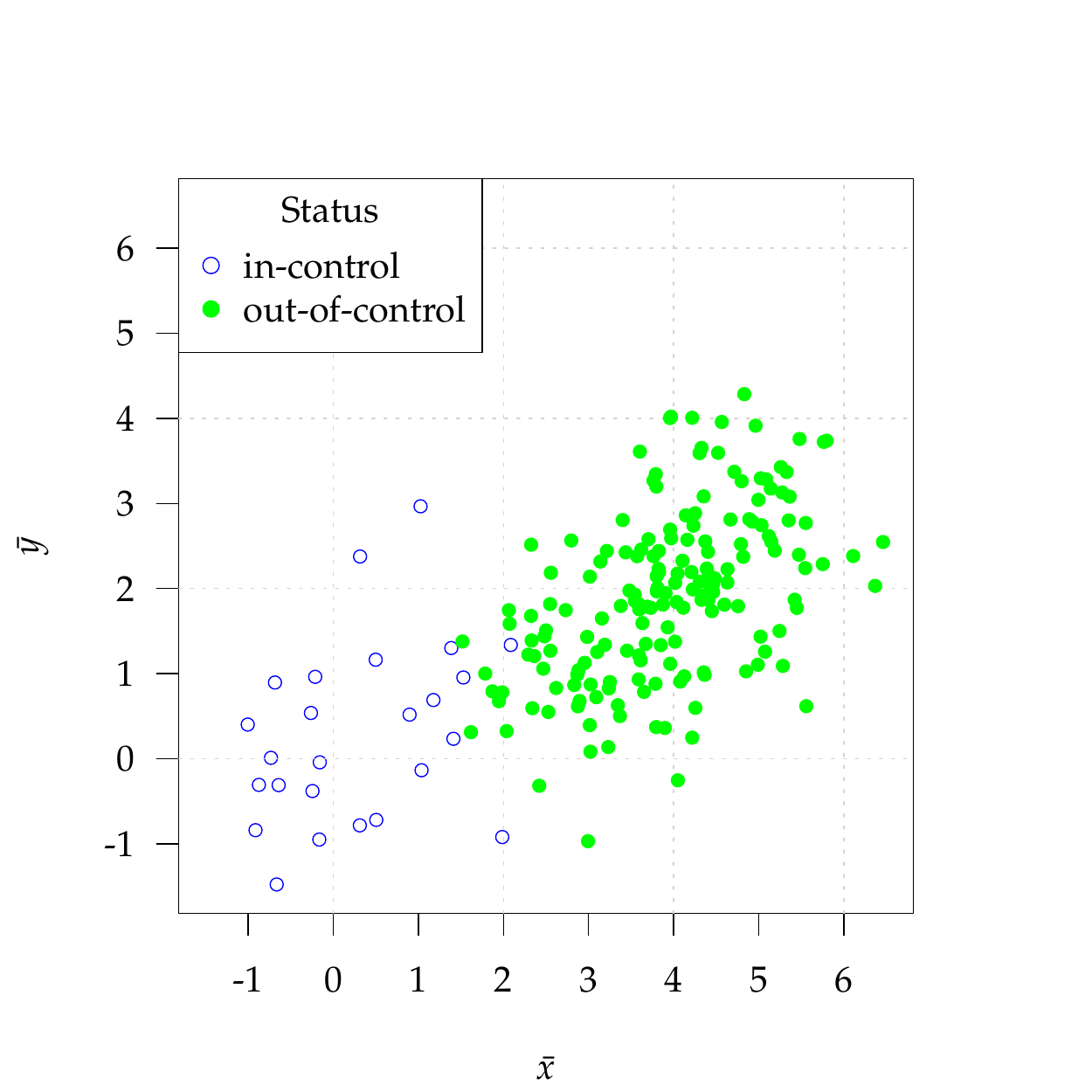}
%\vspace*{-2ex}
\caption{\small Paired ($\bar x, \bar y$) observations used in calculating the AIB-EWMA chart with masked shift (Case II).} \label{fig:02}
\end{figure}

Through simulations, we found that when we generated Phase II observations from
$\mathcal{N}_2(\mu_{Y1}, \mu_{X1}^*, \sigma_Y, \sigma_X, \varrho_{Y,X})$ such that $\mu_{Y1} = \mu_{Y0} + \delta_Y \sigma_Y / \sqrt{n}$
and $\mu_{X1}^* = \mu_{X0} + \delta_Y \sigma_Y / (\beta \sqrt{n})$, 
\begin{figure}[hbt]
\centering
\includegraphics[width=0.65\textwidth]{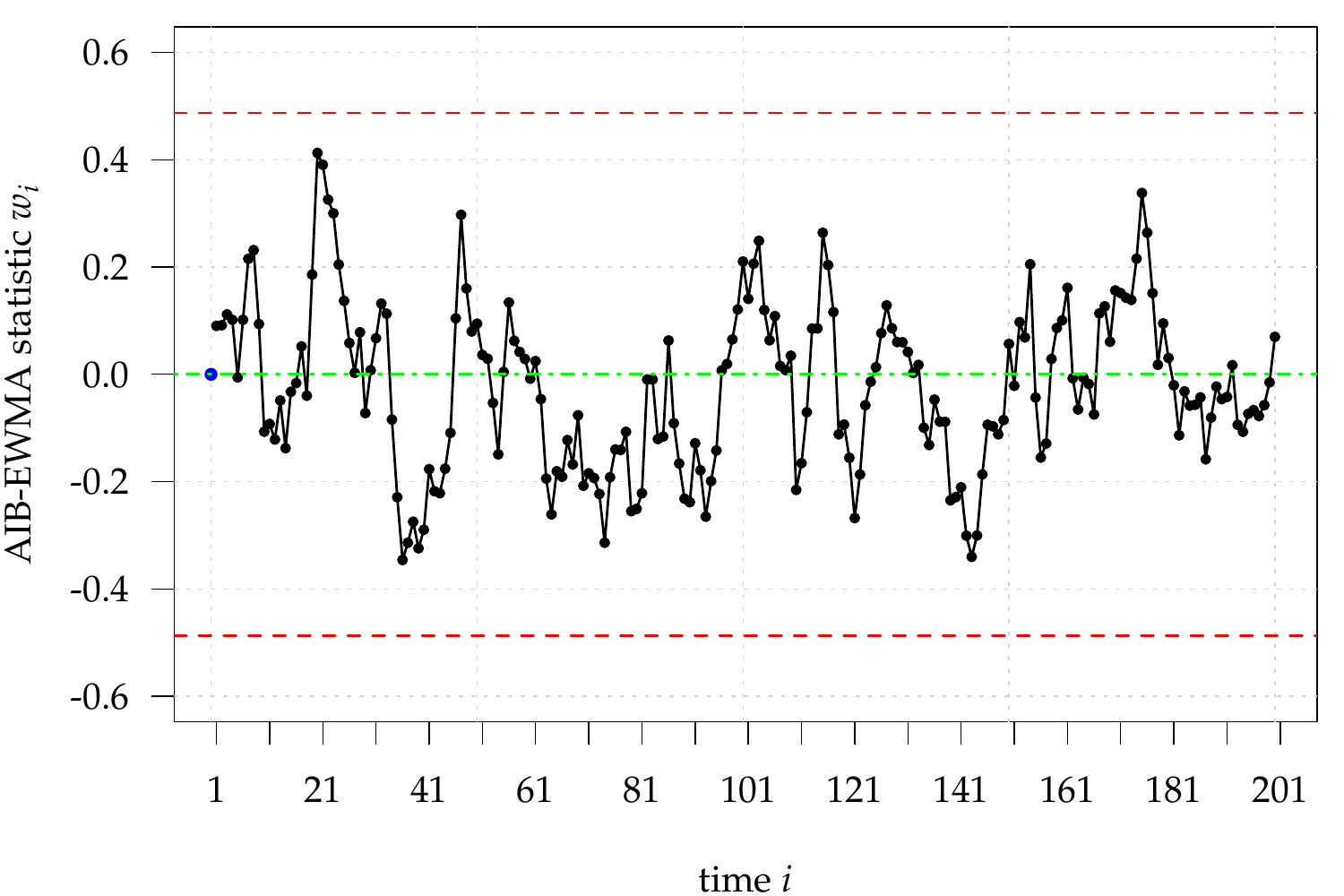}
%\vspace*{-2ex}
\caption{\small An example for an AIB-EWMA chart with a masked shift of size $\delta_Y=2$ (Case II).} \label{fig:01}
\end{figure}
where $\beta = \varrho_{Y,X} \sigma_Y / \sigma_X$, the ARL remains
fixed at 200 (i.\,e. in-control ARL value) for all possible values of $\delta_Y$. Figure~\ref{fig:01} presents an
AIB-EWMA with $\lambda=0.10$ and $L_e=2.454$, for 200 Phase II observations; such that the first 25 observations are in-control and thus are
generated from $\mathcal{N}_2(\mu_{Y0}, \mu_{X0}, \sigma_Y, \sigma_X, \varrho_{Y,X})$, while the rest of observations are out-of-control and
generated from $\mathcal{N}_2(\mu_{Y1}, \mu_{X1}^*, \sigma_Y, \sigma_X, \varrho_{Y,X})$, with $\varrho_{Y,X}=0.5$ and $\delta_Y=2$.
As shown, up to time $t=200$, the chart did not signal the change that was imposed on the mean of $Y$ at the 26$^\text{th}$ sample, although it is of a large magnitude ($\delta_Y=2$). If the auxiliary characteristic mean ($\mu_X$) had been in-control, the chart would have signaled this shift size after on average 3.3 samples. Figure~\ref{fig:02} presents a scatterplot
for the generated 200 paired ($\bar x, \bar y$) observations used in calculating the chart statistic defined in Equation~\eqref{eq:AIBewma}. The blue circles represent the first 25 in-control observations, while the green circles represent the rest of the observations that are out-of-control with $\delta_Y=2$. Even larger shifts along the regression line would be undetectable. 

Generally, we can conclude that a change in the mean of $X$ can delay the detection of out-of-control conditions in $Y$ and, in the worst case, can entirely mask them.

\clearpage

\section{AIB Control Charts and Profile Monitoring} \label{sec:AIBprofile}

In some practical situation, the quality of a process can be best presented by a functional relationship (profile) between a dependent variable and one or more explanatory variables. A profile monitoring technique is commonly used in order to ensure the stability of this relationship over time. A profile can be presented by a simple or multiple, linear or non-linear regression model. We consider here the simple linear profile model; which is defined for samples of size $n$ pairs ($x, y$) drawn over the time $j$ as
\begin{equation*}
  Y_{ij} = A_0 + B_0 X_{ij} + \varepsilon_{ij} \quad,\; i = 1, 2, \ldots n \quad \text{ and } \quad j = 1, 2, \ldots ,
\end{equation*}
where the error terms $\varepsilon_{ij}$ are assumed to be i.i.d. normally distributed with mean of zero and in-control variance of $\sigma_0^2$, and $A_0$ and $B_0$ are the in-control intercept and slope, respectively. It is common to scale the average of the $X$-values to zero to achieve independence between the least squares estimators of $A_0$ and $B_0$.

Monitoring a simple linear profile includes either monitoring the regression parameters simultaneously or monitoring them individually by separate control charts
%(Noorossana et al. (2011)).
\citep{Noor:Sagh:Amir:2011}.
With one of the simultaneous monitoring approaches proposed by
%Kang and Albin (2000)
\cite{Kang:Albi:2000}
one monitors the average deviation from the in-control regression line. A deviation from the in-control line is defined as;
$d_{ij} = Y_{ij} - A_0 - B_0 X_{ij}$, $i = 1, 2, \ldots, n$, and $j = 1, 2, \ldots$, and thus the average
is defined as; $\bar d_j = \bar Y_j - A_0 - B_0 \bar X_j$.

If one considers the commonly used AIB estimator defined in Equation~\eqref{eq:olsE};
and add a constant $A_0$ to both equation sides, we would have
\begin{equation*}
  \hat{\mu}_Y = (\bar y - A_0 + \beta \bar x) + (A_0 + \beta \mu_x) \,.
\end{equation*}
Thus, the AIB estimator in this case represents the average deviation from the regression line with an added constant. Hence, the use of this AIB estimator implies that this AIB charting technique is equivalent to monitoring the average deviation around an assumed regression line, a
technique that was proposed by
%Kang and Albin (2000).
\cite{Kang:Albi:2000}.
However,
%Kang and Albin (2000)
\cite{Kang:Albi:2000}
proposed monitoring this average deviation in conjunction with the variation about the regression line using a range control chart; a feature that is not part of the AIB charts. Additionally, a drawback for monitoring only the average deviation is that a shift in the slope cannot be detected efficiently when the $X$-values are scaled to have an average of zero. We also note that if the $X$-variable is itself monitored, the resulting chart would be equivalent to the cause-selecting chart.

\section{Discussion} \label{sec:discussion}

In our expository paper, we provided an extensive review of the rapidly growing literature on integrating the AIB estimation technique into statistical process monitoring. We warn against the use of AIB monitoring for several reasons. 

The AIB-estimation techniques are useful in enumerative studies where the main interest is to estimate the parameter of a fixed population, i.e., with a fixed sampling frame. This is the context in which they were proposed. Control charts, however, are useful in analytic studies where the samples are produced from a changing, dynamic process at a given time with the main focus on prediction and process improvement.
%Deming (1953, 1993, p. 103)
\citet[p. 103]{Demi:1953, Demi:1993}
emphasized the importance of understanding the difference between enumerative studies and analytical problems. AIB monitoring methods are based on taking enumerative estimation methods and combining them with analytic monitoring methods. In order to do this, very restrictive assumptions are required.

The parameters of the auxiliary characteristic are always assumed to be known and to remain fixed. This is implausible in manufacturing processes unless a parallel monitoring and control scheme is followed on the auxiliary characteristics, something that is never discussed by AIB-chart researchers. Running side-by-side monitoring for the auxiliary variables would bring us back to the already existing multivariate quality control techniques. Under a cascade process property, we would have the cause-selecting methods.
If we want to monitor the relationship between $X$ and $Y$, we would be in the profile monitoring territory. None of these standard
approaches are  adversely affected by a change in the distribution of $X$ and in most cases they can be set up to detect such an occurrence.

We showed that even a relatively small change in the mean of the auxiliary characteristic leads to frequent false alarms, out-of-control alarms while the characteristic of interest is in fact in control. Moreover, a change in the mean of an auxiliary variable can delay the detection of a change in the mean of the characteristic of interest to the extent that it can entirely mask it. 

There are serious pitfalls in the use of the AIB charting techniques. If any parameter of the auxiliary variable changes, statistical performance will be affected. Besides the mean, the slope of any line relating the variable of interest to the auxiliary variable and the correlation level between the variables are assumed to be known and to remain constant. In our view the amount of research on AIB monitoring methods far exceeds their usefulness in practice.

%\newpage

% References

\bibliographystyle{asa}
\bibliography{AIB.bib}

\end{document}